\documentclass[prd,twocolumn,aps,floats, 
floatfix,showpacs,showkeys,amsmath,amssymb,nofootinbib]{revtex4-1}

\usepackage{dcolumn}
\usepackage{bm}
\usepackage{txfonts}
\usepackage{pxfonts}
\usepackage{graphicx}
\usepackage{psfrag}
\usepackage{times}

\begin{document}
\title{Penetrating the horizon of a hydrodynamic white hole}

\author{Nisha Jangid}\email{202121002@dau.ac.in}

\author{Arnab K. Ray}\email{arnab\_kumar@dau.ac.in}
\affiliation{Dhirubhai Ambani University, School of Technology, 
Gandhinagar 382007, Gujarat, India}

\date{\today}

\begin{abstract}
In a shallow-water radial outflow the horizon of a hydrodynamic
white hole coincides with a standing circular hydraulic jump. 
The jump, caused by viscosity, makes the horizon 
visible as a circular front, standing as a barrier against the 
entry of waves within its circumference. 
The blocking of waves causes a pile-up at the horizon of the 
white hole, for which surface tension is mainly
responsible. Conversely, it is also because of surface 
tension that the waves can penetrate the barrier. The 
penetrating waves (analogue Hawking quanta) tunnel through 
the barrier with a decaying amplitude, but a large-amplitude
instability about the horizon is possible. 
\end{abstract}



\maketitle

\section{Introduction} 
\label{sec1}
Quantum mechanical effects make it possible for black holes to 
emit blackbody radiation at a temperature that is inversely
proportional to the mass of the black hole~\citep{swh74,swh75}. 
Due to this inverse dependence, the blackbody radiation emitted by 
massive
astrophysical black holes occurs on temperature scales that are 
far below the temperature of the Cosmic Microwave Background 
Radiation. This makes the detection of blackbody 
radiation from astrophysical black holes, i.e. the  
Hawking radiation, 
a practical impossibility. One then has to turn to analogue
models of gravity to study radiating black holes. Fluid systems
readily provide such analogues because of a mathematical closeness
between the behaviour of fields near black holes and waves in 
transcritical fluid flows~\citep{wgu81,tj91,wgu95,vis98}.  
Indeed, analogues of gravity in a diverse range of physical systems 
have been studied by now (see~\citep{blv11} for a detailed review), 
which gives us to realize that the phenomenon of Hawking radiation 
is not restricted to quantum gravity alone~\citep{vol06,wtpul11}. 

White holes are time-reversed black 
holes~\citep{dme74,swh76,ws80,rs01} and as such fluid analogues 
exist for white holes as well, with analogue horizons suited for  
studying Hawking radiation both theoretically and
experimentally~\citep{su02,rmmpl08,wtpul11,mp14}. While converging 
flows are the usual fluid analogues of black holes~\citep{wgu81}, 
free-surface liquid flows diverging radially from a point  
are viewed 
as convenient fluid analogues of white 
holes~\citep{vol05,rb07,jpmmr,jkb17,br21}.  
Our study here is based on such a fluid system. 

We consider an axially symmetric, radially diverging, shallow 
flow confined to the equatorial plane. The flow originates at a 
point where a vertically downward liquid jet impinges on the plane. 
Thereafter, at a critical radius the 
speed of the radially outflowing liquid equals the local speed of 
capillary-gravity waves~\citep{vol05} or surface gravity 
waves~\citep{rb07}. At this critical radius a barrier is thus formed 
against the upstream transmission of information, effectively 
making the barrier a circular hydrodynamic white hole. The horizon 
of this white hole demarcates a circular boundary that rigidly segregates
a supercritical region inside it from a subcritical region outside.  

The circular horizon can be easily noticed because its 
circumference coincides with a 
standing feature in the flow known as the circular 
hydraulic jump~\citep{tan49}. It is an abrupt discontinuity 
in the free-surface height of the flowing liquid, with the 
post-jump height being greater than the pre-jump height.  
A hydraulic jump forms because of energy dissipation
at the discontinuity, even though momentum and matter flux
conservation are maintained~\citep{jws14,ll87}. 
Jumps with positions of the centimetre order are formed 
because of viscosity in both radial 
flows~\citep{tan49,watson,bdp93} and channel flows~\citep{sbr05}. 
For jump positions of smaller length scales, surface tension
is the main cause~\citep{ba03}.
We also note here that without viscosity 
there is no flow solution in the 
supercritical region~\citep{bdp93}, which renders 
a transcritical flow and an associated 
horizon meaningless. Clearly, 
viscosity and the hydraulic jump cannot be avoided in the 
fluid analogue of a radially diverging free-surface shallow liquid 
flow. Therefore, we take this as our base state, and about it we 
keep surface tension as a small effect. 

In the present work, our objective is to study how the horizon
of a hydrodynamic white hole, coinciding with a circular hydraulic
jump, is tunnelled through (penetrated, to be more general) by analogue 
Hawking quanta because of surface 
tension. Studies on tunnelling have been reported both  
in general relativity~\citep{pw2k,pm07} and in fluid 
analogues of gravity~\citep{vol06,jkb17,akr20}, but the latter 
are not specifically related to the effect of surface tension 
about the analogue horizon. 
In Sec.~\ref{sec2} we set down 
the relevant height-averaged equations of a
shallow-water outflow pertaining
to the standard Type-I hydraulic jump~\citep{bdp93,behh96}. 
In Sec.~\ref{sec3} we show how surface tension, as a small
effect about a viscous steady base flow, perturbatively
shifts the transcritical conditions. 
In Sec.~\ref{sec4} we establish the metric of a hydrodynamic
white hole at the transcritical point of the flow. We also 
discuss how viscosity and gravity scale the jump radius. 
In Sec.~\ref{sec5} we show how surface tension is responsible 
for a pile-up just outside the horizon of a hydrodynamic white
hole, in agreement with a theory about general 
relativistic white holes~\citep{dme74}.  
And in Sec.~\ref{sec6} we show how surface
tension restricts arbitrary blue-shifting of incoming waves
just outside the hydrodynamic horizon and enables analogue 
Hawking quanta to tunnel through with a decaying amplitude.  
We also look at the possibility that the 
penetrating waves may have a growing amplitude, which will 
cause a surface tension-driven instability about the horizon. 

\section{The shallow flow in axial symmetry} 
\label{sec2} 
Circular hydraulic jumps can be created in laboratory experiments 
by impinging a vertically downward jet of liquid (water) on a horizontal 
plane. From the point of impingement  
the liquid flows out radially 
in a thin layer, maintaining an axial symmetry about the downward jet. 
After proceeding up to a certain radius the free-surface 
height of the flowing liquid increases abruptly to form a standing 
circular front. This standing front is commonly known as the circular 
hydraulic jump. 
Circular hydraulic jumps are categorized into Type-I and Type-II
states~\cite{behh96}. In Type-I jumps the flowing liquid 
falls freely off the outer boundary of the horizontal base 
plane~\cite{behh96}, while in Type-II jumps the flow is partially 
restricted at the outer boundary of the base plane~\cite{behh96}. 
Our study here is related to the Type-I circular hydraulic jump. 

The liquid outflow is mathematically framed in the cylindrical 
coordinate system, $(r,\phi,z)$~\citep{ll87}, whose advantage is 
that the axial symmetry of the flow renders it 
independent of the azimuthal coordinate, $\phi$. 
Moreover, with the flow being shallow, a vertical height-averaging 
of the flow variables can be carried out, under the boundary conditions 
that velocities vanish at $z=0$ (the no-slip condition), and vertical
gradients of velocities vanish at the free surface of the flow
(the no-stress condition)~\citep{bdp93,bpw97,sbr05,kas08}. 
The boundary conditions hold true under the assumption 
that the vertical component of the velocity is small 
compared to its radial component, and the vertical variation of 
the radial velocity (through the shallow liquid layer) is much 
greater than its radial variation~\citep{bdp93}.
Accordingly, quantities carrying the $z$-coordinate are 
vertically-averaged through the flow height and  
the double $z$-derivative is approximated as
$\partial^2/\partial z^2 \equiv -1/h^2$~\citep{bdp93}, 
where $h$ is the free-surface height of the shallow flow. 

The local variables of the flow are $h$ and the 
vertically-averaged radial velocity, $v$. Their coupled dynamics
is governed by the continuity equation~\citep{rb07,rsbb18,br21} 
\begin{equation} 
\label{conteq} 
\frac{\partial h}{\partial t} + \frac{1}{r} 
\frac{\partial}{\partial r}
\left(rvh \right) =0 
\end{equation} 
and the radial component of the height-averaged Navier-Stokes 
equation~\citep{rb07,rsbb18,br21}  
\begin{equation} 
\label{avgns} 
\frac{\partial v}{\partial t} + v\frac{\partial v}{\partial r} 
+\frac{1}{\rho}\frac{\partial P}{\partial r}= -\frac{\nu v}{h^2},
\end{equation} 
with $\nu$ being the kinematic viscosity and $P$ the pressure. 
The viscosity-dependent 
term on the right hand side of Eq.~(\ref{avgns}) is the outcome 
of the approximation that 
$\nu \nabla^2 v\simeq -\nu v/h^2$ for a shallow flow~\citep{bdp93}. 
The solutions of Eqs.~(\ref{conteq}) and~(\ref{avgns}), 
$h(r,t)$ and $v(r,t)$, can be known upon prescribing a function 
for $P$ in Eq.~(\ref{avgns}). Contribution to $P$ comes from 
both the hydrostatic effect and surface tension, the latter 
as given by Laplace's formula~\citep{ll87,ba03,dab19}. 
Their total effect together gives     
\begin{equation} 
\label{press} 
P= h \rho g - \frac{\sigma}{r}\frac{\partial}{\partial r} 
\left[\frac{r}{\sqrt{1+(\partial h/\partial r)^2}}
\frac{\partial h}{\partial r}\right]. 
\end{equation} 
The first term on the right hand side of Eq.~(\ref{press}) is the 
hydrostatic pressure, containing the liquid density, $\rho$, and 
the acceleration due to gravity, $g$. The second term on the 
right hand side of Eq.~(\ref{press}) is what 
surface tension, $\sigma$, contributes to the pressure. With $P$ 
expressed in terms of $h$ and $r$, the coupled system consisting 
of Eqs.~(\ref{conteq}) and~(\ref{avgns}) forms a closed set. 

\section{The steady flow conditions}
\label{sec3}
In the steady state the shallow radial flow is free of explicit 
time-dependence, whereby $\partial/\partial t \equiv 0$. This 
condition allows us to integrate the spatial part of
Eq.~(\ref{conteq}) over the full circular front of the flow 
to obtain 
\begin{equation} 
\label{intcon} 
2\pi rvh = Q, 
\end{equation} 
in which $Q$ is the steady volumetric flow rate (a constant of the 
motion). Further, in the steady limit Eq.~(\ref{avgns}) appears as 
\begin{equation} 
\label{ordns} 
v\frac{{\mathrm d}v}{{\mathrm d}r} 
+\frac{1}{\rho}\frac{{\mathrm d}P}{{\mathrm d}r}= -\frac{\nu v}{h^2}. 
\end{equation}
On solving Eqs.~(\ref{intcon}) and~(\ref{ordns}), qualitatively 
different solutions of $h(r)$ and $v(r)$ result, depending 
on the presence and the absence of viscosity, $\nu$, and surface 
tension, $\sigma$, in Eq.~(\ref{ordns}). In what follows, we consider 
these conditions case by case.  
\subsection{Case 1: $\nu = \sigma =0$}
\label{sec3sub1}
Using Eqs.~(\ref{press}),~(\ref{intcon}) and~(\ref{ordns}) in this 
ideal-fluid case, we obtain
\begin{equation} 
\label{dhdr1} 
\frac{{\mathrm d}h}{{\mathrm d}r} = 
\frac{Qv}{2\pi r^2 (gh-v^2)}, 
\end{equation} 
a result that shows the existence of a singularity for $h(r)$ when 
$v^2 = gh$. Noting that $\sqrt{gh}$ is the speed of surface gravity 
waves in the shallow flow~\citep{ll87}, for a radial outflow the 
singularity corresponds to the positive root of $v$, which is
\begin{equation} 
\label{vcrit}
v= \sqrt{gh}.
\end{equation}

When $\nu = \sigma =0$, the integral solution of Eq.~(\ref{ordns}), 
by making use of Eq.~(\ref{press}), is 
\begin{equation} 
\label{bernoul} 
\frac{v^2}{2} + gh = E, 
\end{equation} 
which is the Bernoulli equation with a conserved total 
energy, $E$ (another constant of the motion). 
Using Eqs.~(\ref{intcon}),~(\ref{vcrit}) and~(\ref{bernoul}), 
we find that the singularity for $h(r)$ occurs at the radius, 
\begin{equation} 
\label{radmin} 
r=r_\mathrm{min} = \frac{3\sqrt{3}gQ}{2\pi (2E)^{3/2}}. 
\end{equation}
It is known that no flow solution exists 
for $r < r_\mathrm{min}$~\cite{bdp93}, but two flow solutions
are possible for $r > r_\mathrm{min}$~\cite{bdp93}. On combining
Eqs.~(\ref{intcon}) and~(\ref{bernoul}), 
the existence of the two solutions  can be asymptotically 
verified from
\begin{equation} 
\label{rasymp} 
r= \frac{Q}{2\pi h\sqrt{2(E-gh)}}. 
\end{equation} 
When $r \longrightarrow \infty$ in Eq.~(\ref{rasymp}), the asymptotic
solutions are either 
\begin{equation} 
\label{asymp1} 
h \simeq \frac{E}{g}, 
\quad v \simeq \left(\frac{gQ}{2\pi E}\right)\frac{1}{r}
\end{equation}
or 
\begin{equation} 
\label{asymp2} 
v \simeq \sqrt{2E}, 
\quad h \simeq \left(\frac{Q}{2\pi \sqrt{2E}}\right)\frac{1}{r}. 
\end{equation} 
While Eqs.~(\ref{asymp1}) and~(\ref{asymp2}) produce two distinct 
states
for large $r$ in an ideal fluid~\cite{bdp93}, and one of the states, 
as given by Eq.~(\ref{asymp1}), does agree broadly with experimental 
results~\cite{ot66,behh96}, the circular hydraulic jump itself, as 
a standing boundary between an inner outflow solution and an outer 
outflow solution, does not emerge from the inviscid (non-dissipative) 
theory~\cite{bdp93}. 
\subsection{Case 2: $\nu \neq 0$, $\sigma =0$}
\label{sec3sub2}
As a general principle,  
energy dissipation at the discontinuity creates the circular 
hydraulic jump~\cite{jws14}, and accordingly, an ideal-fluid 
approach will prove inadequate. Hence, we take up a theory that 
accounts for dissipation in the radial outflow. The most obvious means
of dissipation is viscosity, $\nu$, as appears in Eq.~(\ref{ordns}).   
Thus, with $\nu \neq 0$ in Eq.~(\ref{ordns}) and $\sigma =0$ in 
Eq.~(\ref{press}), we derive a first-order equation for $h(r)$ as
\begin{equation} 
\label{dhdr2} 
\left(g-\frac{v^2}{h}\right) \frac{{\mathrm d}h}{{\mathrm d}r} = 
\left(\frac{v^2}{r}- \frac{\nu v}{h^2}\right). 
\end{equation}
Eq.~(\ref{dhdr2}) has a fixed point 
when $v= \sqrt{gh}$ and $r = vh^2/\nu$~\cite{stro}. 
The former condition is the same as in Eq.~(\ref{vcrit})
that gives rise to a singularity in Eq.~(\ref{dhdr1}). Using 
Eq.~(\ref{intcon}), we   
recast Eq.~(\ref{dhdr2}) as a coupled dynamical system
in $h$ and $r$~\cite{tan49,kas08,stro}, 
\begin{equation} 
\label{dyna1} 
\frac{{\mathrm d}h}{{\mathrm d}r} = 
\frac{{\mathrm d}h/{\mathrm d}\tau}{{\mathrm d}r/{\mathrm d}\tau} = 
\frac{f_1(r,h)}{f_2(r,h)} =  
\frac{h-ar^2}{br^3h^3-r}, 
\end{equation} 
in which $\tau$ is a mathematical parameter, $a = \nu (2\pi/Q)$ 
and $b=g (2\pi/Q)^2$. The fixed point 
of the dynamical system in Eq.~(\ref{dyna1}) is found 
from the conditions $f_1 = f_2 =0$~\cite{stro}, whereupon the fixed-point
coordinates, $(r_\star, h_\star)$, will be 
$r_\star = a^{-3/8}b^{-1/8}$ and  $h_\star = (a/b)^{1/4}$. 
In terms of the flow constants, 
\begin{equation} 
\label{rcrit} 
r_\star = (2\pi)^{-5/8} Q^{5/8} \nu^{-3/8} g^{-1/8}, 
\end{equation} 
which agrees with a known scaling relation for the radius of the 
hydraulic jump in the shallow-water approximation~\cite{bdp93}. The 
nature of the fixed point of the dynamical system in Eq.~(\ref{dyna1}) 
is determined from its Jacobian matrix~\cite{stro}. It 
leads to two complex eigenvalues, $\Lambda_{1,2} = (3 \pm i\sqrt{23})/2$, 
indicating that the fixed point, $(r_\star, h_\star)$, is a spiral. 
Mathematical 
solutions of $h(r)$ spiral about the fixed point, making them 
multiple-valued in its immediate neighbourhood~\cite{tan49}. However, 
solutions of a physical fluid flow cannot be multiple-valued. Hence, 
in such situations single-valued inner solutions 
are joined to single-valued outer solutions 
through a standing shock in the vicinity of 
the fixed point (the shock need not pass through the fixed 
point)~\cite{bdp93}. This standing shock is the circular hydraulic
jump in the shallow flow, acting as a discontinuous circular front
between two regions of the radial outflow --- the super-critical 
region where $r < r_\star$ and $v > \sqrt{gh}$, and the sub-critical 
region where $r > r_\star$ and $v < \sqrt{gh}$~\cite{bdp93}. 
Clearly, viscosity establishes a steady inner solution
that connects the origin of the radial outflow to the circular 
hydraulic jump. This solution is absent in the inviscid theory. 
As we shall see in Sec.~\ref{sec4},
viscosity, as a dissipative mechanism~\cite{jws14}, is also 
instrumental in the formation of the hydraulic jump, for which 
the critical condition of $v = \sqrt{gh}$ is not  
enough~\cite{bdp93,jkb17}.   
\subsection{Case 3: $\nu \neq 0$, $\sigma \neq 0$}
\label{sec3sub3}
Viscosity makes it possible for us to get steady single-valued 
solutions that extend
from the origin of the radial outflow to the outer boundary of the 
flow. Between these two spatial limits a discontinuous transition 
occurs from one solution regime to another at the position 
of the circular hydraulic jump. We shall consider this entire set of 
physical conditions as the steady base state in our study hereafter. 
About this base state, we introduce surface tension through 
Eq.~(\ref{press}), and note how in consequence Eq.~(\ref{dhdr2}) 
is changed from the first-order to the third-order as
\begin{multline} 
\label{dhdr3} 
\left(g-\frac{v^2}{h}\right) \frac{{\mathrm d}h}{{\mathrm d}r} 
= \left(\frac{v^2}{r}- \frac{\nu v}{h^2}\right) \\
+ gl^2 \frac{{\mathrm d}}{{\mathrm d}r} \left[\frac{1}{r} 
\frac{{\mathrm d}}{{\mathrm d}r} 
\left\{\frac{r}{\sqrt{1+({\mathrm d}h/{\mathrm d}r)^2}}
\frac{{\mathrm d}h}{{\mathrm d}r}\right\} \right]. 
\end{multline}
In Eq.~(\ref{dhdr3}) surface tension is expressed in terms 
of the capillary length, $l=\sqrt{\sigma/(\rho g)}$~\cite{ll87}, 
so that the effect of surface tension can be scaled against  
any characteristic length scale of the flow system. 

The free-surface height of the flow does not undergo rapid 
variations, except around the hydraulic jump. 
Hence, away from the hydraulic jump, with small spatial 
gradients of $h$, i.e. ${\mathrm d}h/{\mathrm d}r \simeq 0$, 
the surface tension term in Eq.~(\ref{dhdr3})
does not have much of an impact on the steady base flow, as 
derived from Eq.~(\ref{dhdr2}).  
We may then address only the question of how surface tension affects 
the hydraulic jump, where the free-surface height increases 
noticeably over a small radial distance. The answer will be known 
to a certain extent 
from the corrections that surface tension makes to the critical 
jump conditions, as given by Eqs.~(\ref{vcrit}) and~(\ref{rcrit}).  

Adopting a heuristic approach to finding these corrections, 
we ignore all
spatial derivatives of $h(r)$ that are higher than that of 
the first order in Eq.~(\ref{dhdr3}). This approximation will 
be all the more reasonable if the flow profile at the jump does 
not have a large curvature. 
Moreover, with ${\mathrm d}h/{\mathrm d}r$ being ${\mathcal O}(1)$ 
around the jump radius, we approximate 
$\sqrt{1 + ({\mathrm d}h/{\mathrm d}r)^2} \sim 1$. Following all 
of this we get 
\begin{equation} 
\label{dhdr4} 
\left(g-\frac{v^2}{h} +\frac{gl^2}{r^2} \right) 
\frac{{\mathrm d}h}{{\mathrm d}r} \simeq 
\left(\frac{v^2}{r}- \frac{\nu v}{h^2}\right). 
\end{equation}
To know the fixed point of Eq.~(\ref{dhdr4}), we apply the same 
arguments that follow Eq.~(\ref{dhdr2}). This gives  
$v^2 \simeq gh_\star + gh_\star (l^2/r_\star^2)$, in which we 
have taken $h \simeq h_\star$, an approximation that is consistent 
with ignoring any spatial derivative of $h$ that is higher than 
the first. We have also taken $r \simeq r_\star$ in the term that 
contains $l$. By this approach overall, we 
treat surface tension as a small perturbative effect around 
the results given in Eqs.~(\ref{vcrit}) and~(\ref{rcrit}). Now, 
for water, $l=0.27\,\mathrm{cm}$ and $r_\star \lesssim 10\,\mathrm{cm}$, 
which implies that $l^2/r_\star^2 \sim 10^{-4} \ll 1$. This smallness 
allows a binomial expansion to the lowest order in $l^2/r_\star^2$
and modifies the fixed point of $v$,  
shifted perturbatively by surface tension, as 
\begin{equation} 
\label{veesig} 
v_{\star \sigma} \simeq 
\sqrt{gh_\star}\left(1+\frac{l^2}{2r_\star^2}\right). 
\end{equation} 
Likewise, the right hand side of Eq.~(\ref{dhdr4}) gives the fixed
point of $r$, shifted slightly due to surface tension, as 
$r_{\star \sigma} \simeq v_{\star \sigma} h_\star^2/\nu$. 
The effect of surface tension in $r_{\star \sigma}$ is captured 
through $v_{\star \sigma}$, as given by Eq.~(\ref{veesig}). The 
full expression for $r_{\star \sigma}$ will thus be 
\begin{equation} 
\label{rsig}
r_{\star \sigma} \simeq r_\star \left(1+\frac{l^2}{2r_\star^2}\right).
\end{equation} 
What we see in both Eqs.~(\ref{veesig}) and~(\ref{rsig}) is that 
the fractional shift of the fixed point values of $v$ and $r$, 
caused by surface tension, 
is ${\mathcal O} (l^2/r_\star^2)$. Since $l \ll r_\star$, this is 
a small shift with respect to the fixed points of the steady base 
flow that is generated by Eq.~(\ref{dhdr2}) for $\nu \neq 0$ but 
$\sigma =0$. The smallness of the shift is not just consistent with 
our heuristic approach that leads to Eqs.~(\ref{veesig}) 
and~(\ref{rsig}), 
but is also consistent with the observation that 
surface tension has a small influence on the radius of a hydraulic
jump in laboratory settings~\cite{ba03}.  

\section{A hydrodynamic white hole}
\label{sec4}
The vertically-averaged radial outflow is governed by the 
coupled variables, $v(r,t)$ and $h(r,t)$. 
About their steady solutions, $v_0(r)$ and $h_0(r)$, we apply 
time-dependent perturbations, $v^{\prime}(r,t)$ and $h^{\prime}(r,t)$,
respectively. This gives us 
$v(r,t)=v_0(r)+v^{\prime}(r,t)$ and $h(r,t)=h_0(r)+h^{\prime}(r,t)$.
Going by the form of Eq.~(\ref{conteq}) now we devise an Eulerian
perturbation scheme with a variable, $f(r,t)=rvh$. Under steady 
conditions
this becomes $f=f_0=rv_0h_0=Q/2\pi$, a constant, as Eq.~(\ref{intcon}) 
shows. Perturbing with $f^{\prime}(r,t)$ about $f_0$, we write 
$f(r,t)=f_0+f^{\prime}(r,t)$, from which, on linearizing in $v^\prime$
and $h^\prime$, we get 
\begin{equation}
\label{radf} 
f^\prime =r\left(v_0h^\prime + h_0 v^\prime \right). 
\end{equation} 
Now applying Eq.~(\ref{radf}) to Eq.~(\ref{conteq}), we derive 
a linear relation between $h^\prime$ and $f^\prime$ as 
\begin{equation}
\label{rconf} 
\frac{\partial h^\prime}{\partial t} = -\frac{1}{r} 
\frac{\partial f^\prime}{\partial r}, 
\end{equation} 
and then applying Eq.~(\ref{rconf}) to Eq.~(\ref{radf}), we derive 
a linear relation between $v^\prime$ and $f^\prime$ as 
\begin{equation} 
\label{effvee} 
\frac{\partial v^\prime}{\partial t} = \frac{v_0}{f_0} 
\left(\frac{\partial f^\prime}{\partial t} +v_0 
\frac{\partial f^\prime}{\partial r}\right).
\end{equation} 

In Eq.~(\ref{avgns}) $v$ and $h$ are perturbed likewise to a linear 
order about their steady values. Taking the time derivative 
of the linearized equation that follows, and applying both 
Eqs.~(\ref{rconf}) and~(\ref{effvee}) to it, we get a wave equation,
\begin{multline}
\label{rmetric} 
\frac{\partial}{\partial t}\left(v_0
\frac{\partial f^\prime}{\partial t}\right)
+\frac{\partial}{\partial t}\left(v_0^2
\frac{\partial f^\prime}{\partial r}\right)
+\frac{\partial}{\partial r}\left(v_0^2
\frac{\partial f^\prime}{\partial t}\right) \\
+\frac{\partial}{\partial r}\left[v_0
\left(v_0^2-gh_0\right)
\frac{\partial f^\prime}{\partial r}\right] 
=-\frac{\nu v_0}{h_0^2}\left(\frac{\partial f^\prime}{\partial t} 
+3v_0\frac{\partial f^\prime}{\partial r}\right) \\
-l^2gf_0 \frac{\partial}{\partial r}\left[\frac{1}{r}
\frac{\partial}{\partial r}\left\{
\frac{r}{[1+({\mathrm d}h_0/{\mathrm d}r)^2]^{3/2}}
\frac{\partial}{\partial r}\left(\frac{1}{r}
\frac{\partial f^\prime}{\partial r}\right)\right\}\right]. 
\end{multline}

If $\nu = \sigma=0$, Eq.~(\ref{rmetric}) is compactly 
rendered as
\begin{equation} 
\label{wavecomp} 
\partial_\alpha \left( {\mathsf{f}}^{\alpha \beta}
\partial_\beta f^{\prime}\right) =0,
\end{equation} 
in which the Greek indices run from $0$ to $1$, with
$0$ implying $t$ and $1$ implying $r$. 
From the terms on the left hand side of Eq.~(\ref{rmetric})
we set down the matrix,  
\begin{equation}
\label{symmat}
{\mathsf{f}}^{\alpha \beta } = v_0
\begin{bmatrix}
1 & v_0 \\
v_0 & v_0^2 - gh_0
\end{bmatrix} \\. 
\end{equation}
A hydrodynamic metric and an analogue horizon are based on an 
equivalence between Eqs.~(\ref{wavecomp}) and~(\ref{symmat}) on 
the one hand and the d'Alembertian for a scalar field in curved 
geometry on the other~\citep{rb07} (also see~\citep{blv11} 
and all relevant references therein).
The d'Alembertian has the form~\citep{blv11}
\begin{equation}
\label{alem}
\triangle \psi \equiv \frac{1}{\sqrt{-\mathsf{g}}}
\partial_\alpha \left({\sqrt{-\mathsf{g}}}\, {\mathsf{g}}^{\alpha \beta} 
\partial_\beta \psi \right).
\end{equation}
Identifying ${\mathsf{f}}^{\alpha \beta } =
\sqrt{-\mathsf{g}}\, {\mathsf{g}}^{\alpha \beta}$ and 
$\mathsf{g} = \det \left({\mathsf{f}}^{\alpha \beta }\right)$ 
establishes the horizon of a hydrodynamic white hole for the 
waves when $v_0^2 = gh_0$~\citep{su02,rb07}, a condition 
that also agrees with Eq.~(\ref{vcrit}) and the fixed point 
of $v$ in Eq.~(\ref{dhdr2}).

We note, however, that the horizon of the hydrodynamic 
white hole has been obtained by disregarding viscosity and 
surface tension ($\nu =\sigma =0$) in Eq.~(\ref{rmetric}). Under these 
conditions we may have preserved the symmetry of the metric implied 
by Eqs.~(\ref{wavecomp}) and~(\ref{symmat}), but, as discussed 
in Sec.~\ref{sec3sub2}, without 
viscosity a flow solution within the radius of the hydraulic 
jump will not be physically realizable~\citep{bdp93}. Hence, we have 
to account for viscosity in the flow of a normal liquid, 
even though it will compromise the condition of the analogue horizon. 
For all that, the basic
properties of surface waves will not be affected much~\citep{su02}, 
and the most crucial feature of the 
white-hole horizon will remain qualitatively unchanged, which is that
a wave propagating upstream in the subcritical flow region  
(where $v_0 < \sqrt{gh_0}$) cannot pass through the horizon into 
the supercritical flow region (where $v_0 > \sqrt{gh_0}$), both in 
the presence of viscosity~\citep{rb07} and surface tension~\citep{vol05}. 


Some relevant aspects of the wave equation in Eq.~(\ref{rmetric})  
stand out clearly through a dispersion relation.
With respect to the steady background flow, Eq.~(\ref{rmetric}) becomes
\begin{multline} 
\label{waveq} 
\frac{\partial^2 f^\prime}{\partial t^2}=gh_0 
\frac{\partial^2 f^\prime}{\partial r^2}
-\frac{\nu}{h_0^2}\frac{\partial f^\prime}{\partial t} \\
-l^2gh_0
\left(\frac{\partial^4 f^\prime}{\partial r^4}
-\frac{2}{r}\frac{\partial^3 f^\prime}{\partial r^3}
+\frac{3}{r^2}\frac{\partial^2 f^\prime}{\partial r^2}
-\frac{3}{r^3}\frac{\partial f^\prime}{\partial r}\right),
\end{multline} 
which, when $\nu = \sigma =0$, can be identified as the wave equation 
for gravity waves. 
A solution, $f^\prime (r,t) \sim \exp [i(kr-\omega t)]$, applied to 
Eq.~(\ref{waveq}), gives a quadratic equation in $\omega$ as 
\begin{multline} 
\label{disquad} 
\left(\omega -kv_{\mathrm B}\right)^2
+\frac{i\nu}{h_0^2}\left(\omega -kv_{\mathrm B}\right)
- gh_0 \left(1-\frac{3l^2}{r^2}+l^2k^2\right)k^2 \\
- igh_0\frac{l^2}{r^2}\left(2kr-\frac{3}{kr}\right)k^2 =0,
\end{multline} 
in which $kv_{\mathrm B}$ is due to the bulk motion of the fluid.
The two roots of Eq.~(\ref{disquad}) will have the form,  
$\left(\omega -kv_{\mathrm B}\right)= -i[\nu/(2h_0^2)] 
\pm \left(X+iY\right)$, with $X$ and $Y$ being 
real~\citep{br21}. We are interested in the real part because 
it contributes to the phase of the wave solution and will thus set 
forth the wave velocity. Moreover, 
since we are mainly concerned with how surface 
tension affects the velocity of the waves, we ignore viscosity 
in the real part of the solution of Eq.~(\ref{disquad}) and extract
\begin{equation} 
\label{disper0} 
\omega \simeq kv_{\mathrm B} \pm k \sqrt{gh_0}
\left(1-\frac{3l^2}{r^2}+l^2k^2\right)^{1/2}.
\end{equation}
Carrying out a binomial expansion of Eq.~(\ref{disper0}) in the 
regime of $kl \ll 1$, we get both the phase velocity, $v_{\mathrm p}$, 
and the group velocity, $v_{\mathrm g}$, corrected by surface
tension to ${\mathcal O} (l^2/r_\star^2)$, as 
\begin{equation} 
\label{phasegrav} 
v_{\mathrm p} = \frac{\omega}{k} 
= v_{\mathrm g} = \frac{\partial \omega}{\partial k} 
\simeq v_{\mathrm B} \pm \sqrt{gh_0} 
\left(1 - \frac{3l^2}{2r^2} + \ldots \right), 
\end{equation} 
in broad qualitative similarity with Eq.~(\ref{veesig}). 

In deriving Eq.~(\ref{phasegrav}) we considered 
the regime of $kl \ll 1$. Its relevance in 
our study can be understood 
now by neglecting $l^2/r^2$ in the comoving dispersion relation 
($v_{\mathrm B}=0$) implied by Eq.~(\ref{disper0}). Choosing 
the positive sign, this leads to  
\begin{equation} 
\label{disper} 
\omega \simeq k \sqrt{gh_0} \left(1+l^2k^2\right)^{1/2}, 
\end{equation} 
which is the long-wavelength limit
of the dispersion relation for capillary-gravity waves,   
$\omega^2 = \left[gk+\left(\sigma/\rho\right)k^3\right]
\tanh\left(kh_0\right)$, when $kh_0 \ll 1$~\citep{ll87}. 
In this long-wavelength limit the wavelength, 
$\lambda \gg h_0$, as happens in shallow 
flows~\citep{ll87}. This condition is thus implicit  
in Eq.~(\ref{disper}) and also in all the equations that lead to it, 
starting with Eqs.~(\ref{conteq}) and~(\ref{avgns}). 

Using Eq.~(\ref{disper}) we can derive the scaling 
relation for the radius of the circular hydraulic jump in 
Eq.~(\ref{rcrit}). Since $l < h_0$, for $kl \ll 1$, which also 
implies that $\lambda \gg l$, Eq.~(\ref{disper}) gives the phase 
velocity of gravity waves as 
$v_{\mathrm p} =\omega/k \simeq \sqrt{gh_0}$. 
Now by comparing the first term on the left hand side of 
Eq.~(\ref{avgns}) with the viscosity-dependent term on the 
right hand side, we note that the time scale on which viscosity 
decelerates the outward flow is 
$t_{\mathrm{visc}}\sim h_0^2/\nu$. Information about the  
deceleration of an advanced layer of the flow by viscosity 
will be carried upstream only by surface gravity waves 
travelling against the flow with the speed, $\sqrt{gh_0}$.
Therefore, the downstream deceleration 
will not be known in the supercritical region of the flow, 
where $v_0 > \sqrt{gh_0}$. Thus, the flow in this region 
will proceed radially outwards without any impediment   
till $v_0$ becomes equal to $\sqrt{gh_0}$, and 
only then will information about an obstacle ahead 
catch up with the outflowing fluid. By defining a dynamic time scale
for the bulk motion, $t_{\mathrm{dyn}}\sim r/v_0$, and setting 
$t_\mathrm{visc} \simeq t_\mathrm{dyn}$, along with the conditions, 
$v_0 \simeq v_{\mathrm p} \simeq \sqrt{gh_0}$ and 
$rv_0h_0=Q/2\pi$, we can scale the jump radius, 
$r_\mathrm{J} \sim Q^{5/8} \nu^{-3/8} g^{-1/8}$,
a familiar scaling formula~\citep{bdp93} that 
we also know from Eq.~(\ref{rcrit}). 
Thus, the circular hydraulic jump forms when the two time scales, 
$t_{\mathrm{visc}}$ and $t_{\mathrm{dyn}}$, match each other, and when 
$v_0 = \sqrt{gh_0}$. The combined physical effect of 
these conditions is that a layer of fluid arriving late is halted
at an obstacle created by a layer of fluid ahead, 
slowed abruptly by viscosity. However, the outflowing  
fluid cannot accumulate indefinitely, and flow continuity must 
be maintained. Therefore, the fluid layer arriving late will jump
over the slowly flowing layer ahead and cause a sudden increase 
in the flow height. This is the hydraulic jump~\citep{rb07}, and 
it stands as a discontinuous feature in the single-valued solutions 
of $h_0(r)$ and $v_0(r)$ that connect the origin of the flow to the 
outer boundary. Since the jump is formed where $v_0 = \sqrt{gh_0}$, 
from the perspective of fluid analogues of gravity it behaves like 
the event horizon of a hydrodynamic white hole that blocks the 
propagation of information from the subcritical region of the outflow 
to the 
supercritical region. However, the jump will not be formed only by 
satisfying the analogue horizon condition~\citep{jkb17}. The actual 
physical means by which the jump is created at the horizon is viscosity. 

\section{Pile-up at the horizon}
\label{sec5}
Expanding Eq.~(\ref{rmetric}) with all the derivatives of $f^\prime$, 
we get 
\begin{multline} 
\label{difeqgam1}
\frac{\partial^2 f^\prime}{\partial t^2}+
2\frac{\partial v_0}{\partial r} 
\frac{\partial f^\prime}{\partial t}+ 
2v_0\frac{\partial}{\partial r} 
\left(\frac{\partial f^\prime}{\partial t}\right)+
\left(v_0^2-gh_0\right)
\frac{\partial^2 f^\prime}{\partial r^2} \\
+ \frac{1}{v_0} 
\frac{\partial}{\partial r}\left[v_0\left(v_0^2-gh_0\right)\right]
\frac{\partial f^\prime}{\partial r} 
+ \frac{\nu}{h_0^2}
\left(\frac{\partial f^\prime}{\partial t} 
+3v_0\frac{\partial f^\prime}{\partial r}\right) \\
= -\frac{l^2gf_0}{v_0}
\Bigg[\frac{\Gamma_1}{r}\frac{\partial^4 f^\prime}{\partial r^4}
+\left\{\frac{\partial}{\partial r}\left(\frac{\Gamma_1}{r}\right)
-\frac{\Gamma_1 \Gamma_2}{r^2}\right\} 
\frac{\partial^3 f^\prime}{\partial r^3} \\
+\left\{-\frac{\partial}{\partial r}
\left(\frac{\Gamma_1 \Gamma_2}{r^2}\right)
+\frac{\Gamma_1 \Gamma_2}{r^3}\right\} 
\frac{\partial^2 f^\prime}{\partial r^2}
+ \frac{\partial}{\partial r}\left(\frac{\Gamma_1 \Gamma_2}{r^3}\right)
\frac{\partial f^\prime}{\partial r}\Bigg], 
\end{multline}
in which $\Gamma_1$ and $\Gamma_2$ are to be read, respectively, as
\begin{equation}
\label{defgam} 
\Gamma_1 =\left[1+({\mathrm d}h_0/{\mathrm d}r)^2\right]^{-3/2},\, 
\Gamma_2 = 1+\left[\frac{3r\left({\mathrm d}h_0/{\mathrm d}r\right)}
{1+\left({\mathrm d}h_0/{\mathrm d}r\right)^2}\right]
\frac{{\mathrm d}^2h_0}{{\mathrm d}r^2}. 
\end{equation}
In Eq.~(\ref{difeqgam1}), the viscosity-dependent terms have been 
taken to the left hand side, but the surface tension-dependent terms 
are on the right hand side. What remains in the absence of surface 
tension is a perturbative condition about the steady base state, 
as emerges from Eq.~(\ref{dhdr2}). We now look into how this condition 
is influenced by surface tension, considered as a small effect. 
We treat the perturbation as a
high-frequency travelling wave, whose wavelength, $\lambda$,
does not exceed the radius of the hydrodynamic 
horizon, $r_\star$~\citep{pw2k}. 
However, at the same time, for treating surface tension as a small effect, 
we require $l \ll \lambda$. With these restrictions on $\lambda$, i.e.
$l \ll \lambda < r_\star$, we prescribe 
a solution for the travelling wave as
\begin{equation} 
\label{travsol} 
f^\prime (r,t) =\exp \left[is(r)-i\omega t \right],
\end{equation} 
under the understanding that $\omega$ is much greater than any 
characteristic frequency in the system.
The travelling-wave solution in Eq.~(\ref{travsol}), when applied to 
Eq.~(\ref{difeqgam1}), delivers
\begin{multline}
\label{peewkb}
\left(v_0^2-gh_0 \right)
\left[i\frac{{\mathrm d}^2 s}{{\mathrm d}r^2}
-\left(\frac{{\mathrm d}s}{{\mathrm d}r}\right)^2\right]
+\frac{i}{v_0}
\frac{{\mathrm d}}{{\mathrm d}r}
\left[v_0 \left(v_0^2 -gh_0\right)\right]
\frac{{\mathrm d}s}{{\mathrm d}r} \\
+2v_0\omega \frac{{\mathrm d}s}{{\mathrm d}r}
-2i\omega \frac{{\mathrm d}v_0}{{\mathrm d}r} -\omega^2 
+ \frac{\nu}{h_0^2}\left(-i\omega + 
3iv_0 \frac{{\mathrm d}s}{{\mathrm d}r}\right) \\
= -\frac{l^2gf_0}{v_0}\Bigg[\frac{\Gamma_1}{r} 
\Bigg\{\left(\frac{{\mathrm d}s}{{\mathrm d}r}\right)^4 
-3\left(\frac{{\mathrm d^2}s}{{\mathrm d}r^2}\right)^2 
-4\frac{{\mathrm d}^3 s}{{\mathrm d}r^3}\frac{{\mathrm d}s}{{\mathrm d}r}
-6i\frac{{\mathrm d}^2 s}{{\mathrm d}r^2}
\left(\frac{{\mathrm d}s}{{\mathrm d}r}\right)^2 \\
+i\frac{{\mathrm d}^4 s}{{\mathrm d}r^4}\Bigg\} 
+\Bigg\{\frac{{\mathrm d}}{{\mathrm d}r}\left(\frac{\Gamma_1}{r}\right)
-\frac{\Gamma_1 \Gamma_2}{r^2}\Bigg\} \times 
\Bigg\{
-3\frac{{\mathrm d}^2 s}{{\mathrm d}r^2}\frac{{\mathrm d}s}{{\mathrm d}r}
-i\left(\frac{{\mathrm d}s}{{\mathrm d}r}\right)^3
+i\frac{{\mathrm d}^3 s}{{\mathrm d}r^3} \Bigg\} \\
+\Bigg\{\frac{{\mathrm d}}{{\mathrm d}r}
\left(\frac{\Gamma_1 \Gamma_2}{r^2}\right)
-\frac{\Gamma_1 \Gamma_2}{r^3}\Bigg\} \times 
\Bigg\{\left(\frac{{\mathrm d}s}{{\mathrm d}r}\right)^2
-i\frac{{\mathrm d}^2 s}{{\mathrm d}r^2} \Bigg\} 
+i\frac{{\mathrm d}}{{\mathrm d}r}
\left(\frac{\Gamma_1 \Gamma_2}{r^2}\right)
\frac{{\mathrm d}s}{{\mathrm d}r}\Bigg]. 
\end{multline}
As a solution of Eq.~(\ref{peewkb}), $s(r)$ will 
have both real and imaginary parts. 
Accordingly, we prescribe $s(r)=\alpha (r)+i\beta (r)$,
with both $\alpha$ and $\beta$ being real. From the form 
of $f^\prime$ in Eq.~(\ref{travsol}), we note that while $\alpha$ 
contributes to the phase of the perturbation, $\beta$ contributes 
to its amplitude. Solutions of both $\alpha$ and $\beta$ are to be 
found by a {\it WKB} analysis of Eq.~(\ref{peewkb}), which necessitates
$\alpha \gg \beta$ for travelling waves of high frequency.

In Eq.~(\ref{peewkb}), the highest derivative of $s$
is of the fourth order. This term, however, is
dependent on the surface tension, whose effect in our analysis is 
considered to be feeble. In applying the {\it WKB} approximation
to Eq.~(\ref{peewkb}), we, therefore, adopt an iterative approach. 
We first set $l =0$ on the right hand side of 
Eq.~(\ref{peewkb}), and then solve a second-order differential 
equation in $s(r)$.
For this special case, $s(r)$ is also modified 
as $s_0(r)= \alpha_0 (r)+i\beta_0 (r)$, with the subscript ``$0$"
denoting solutions in the absence of surface tension.
Using $s_0(r)$ in Eq.~(\ref{peewkb}), we
separate the real and the imaginary parts first, and then
set both equal to zero. The {\it WKB} prescription 
stipulates that $\alpha_0 \gg \beta_0$. Hence, we collect the 
real terms without $\beta_0$, and from the resulting quadratic equation 
in ${\mathrm d}\alpha_0/{\mathrm d}r$ we get
\begin{equation} 
\label{kayminus} 
\alpha_0 = \int \frac{\omega}{v_0 \mp \sqrt{gh_0}}\,{\mathrm d}r.
\end{equation}
Likewise, from the imaginary part, on using Eq.~(\ref{kayminus}), 
we get
\begin{equation} 
\label{kaynot} 
\beta_0 = \frac{1}{2} \ln \left(v_0\sqrt{gh_0}\right)  
\mp \int \frac{\nu}{2h_0^2\sqrt{gh_0}}\left(1 -  
\frac{3v_0}{v_0 \mp \sqrt{gh_0}}\right) \,{\mathrm d}r +c_1,
\end{equation} 
with $c_1$ being an integration constant. 

We perform a self-consistency check that $\alpha_0 \gg \beta_0$,
as a basic requirement of the {\it WKB} analysis. First, we note
$\alpha_0$ contains $\omega$ (the high frequency of the travelling
wave), and in this respect is of a
leading order over $\beta_0$, which contains $\omega^0$. Next,
for $r \gtrsim r_\star$ 
(the subcritical region
of the shallow flow, which is of interest to us), 
where $v_0 \sim r^{-1}$ and $h_0 \sim \mathrm{constant}$,     
we get $\alpha_0 \sim \omega r$
from Eq.~(\ref{kayminus}) and $\beta_0 \sim \ln r$ 
from Eq.~(\ref{kaynot}). The contribution of the viscosity term 
in Eq.~(\ref{kaynot}) is even weaker, because for a liquid like 
water, $\nu/(h_0\sqrt{gh_0}) \sim 10^{-3}$. 
Considering all of these facts together, we see that our solution 
scheme is well in conformity with the {\it WKB} prescription.

Now we take up Eq.~(\ref{peewkb}) with $l \neq 0$, whose solution 
is $s = \alpha +i\beta$. However, in keeping with our iterative 
approach we approximate $s \simeq s_0$ on the right hand side of 
Eq.~(\ref{peewkb}). 
Further, since we have also seen that $\alpha_0 \gg \beta_0$,
we approximate $s \simeq s_0 \simeq \alpha_0$ in all the terms 
with $l$ in Eq.~(\ref{peewkb}). Then we note that
the most dominant $\alpha_0$-dependent real term on the right hand 
side of Eq.~(\ref{peewkb}) is of the fourth degree. 
Retaining only this term 
and extracting the $\beta$-independent real terms from the
left hand side, we get a quadratic equation in
${\mathrm d}\alpha/{\mathrm d}r$, 
\begin{equation} 
\label{quadkayminus}
\left(v_0^2 -gh_0 \right)
\left(\frac{{\mathrm d}\alpha}{{\mathrm d}r}\right)^2
-2v_0 \omega \frac{{\mathrm d}\alpha}{{\mathrm d}r} + \omega^2 
\simeq l^2 gh_0 \Gamma_1 
\left(\frac{{\mathrm d}\alpha_0}{{\mathrm d}r}\right)^4.
\end{equation} 
The iterative process behind the derivation of Eq.~(\ref{quadkayminus}) 
is valid for $l \ll \lambda$. Thus, with a binomial approximation for 
small $l$ in the discriminant, the solution of 
Eq.~(\ref{quadkayminus}) finally leads to  
\begin{equation} 
\label{kayminuscor} 
\alpha \simeq \int \frac{\omega}{v_0 \mp \sqrt{gh_0}}\,{\mathrm d}r \pm 
\int
\frac{l^2 \omega^3 \Gamma_1 \sqrt{gh_0}}{2(v_0 \mp \sqrt{gh_0})^4}\,
{\mathrm d}r,
\end{equation} 
in a form that we read as $\alpha= \alpha_0 + \alpha_l$.
The second term on the right hand side of Eq.~(\ref{kayminuscor}) 
adds a surface tension-dependent correction, $\alpha_l$, to $\alpha_0$.  
The order of this correction is $\omega^3$, and apparently 
dominates $\alpha_0$ in the high-frequency regime. 
However, noting that the wavelength, 
$\lambda (r)=2\pi(v_0 \mp \sqrt{gh_0})/\omega$, we realize 
that the correction term in Eq.~(\ref{kayminuscor}) is subdominant 
to $\alpha_0$, when $l \ll \lambda$. The smallness of the correction
validates our iterative method self-consistently.  

Next, to determine $\beta$, we extract all the imaginary
terms from the left hand side of Eq.~(\ref{peewkb}), and we note
that the most dominant contribution to the imaginary terms on
the right hand side is of the third degree in $s$. We approximate 
$s \simeq s_0 = \alpha_0 +i\beta_0$ on the right hand side of 
Eq.~(\ref{peewkb}), as we have done to derive Eq.~(\ref{quadkayminus}). 
With all of this, we get 
\begin{multline} 
\label{quadkaynot} 
2\left[v_0 \omega -\left(v_0^2 -gh_0\right)
\frac{{\mathrm d}\alpha}{{\mathrm d}r}\right]
\frac{{\mathrm d}\beta}{{\mathrm d}r} 
+\frac{1}{v_0}
\frac{\mathrm d}{{\mathrm d}r}
\left[v_0 \left(v_0^2 -gh_0\right)
\frac{{\mathrm d}\alpha}{{\mathrm d}r}\right] \\
-2\omega\frac{{\mathrm d}v_0}{{\mathrm d}r} - \frac{\nu \omega}{h_0^2} 
+ \frac{3\nu v_0}{h_0^2}\frac{{\mathrm d}\alpha}{{\mathrm d}r}  
\simeq l^2gh_0
\left(\frac{{\mathrm d}\alpha_0}{{\mathrm d}r}\right)^3 \\
\times \Bigg[-4\Gamma_1 \frac{{\mathrm d}\beta_0}{{\mathrm d}r} 
+6 \Gamma_1 \frac{{\mathrm d^2}\alpha_0}{{\mathrm d}r^2} 
\left(\frac{{\mathrm d}\alpha_0}{{\mathrm d}r}\right)^{-1}
+r\frac{{\mathrm d}}{{\mathrm d}r}
\left(\frac{\Gamma_1}{r}\right)
-\frac{\Gamma_1 \Gamma_2}{r} \Bigg]. 
\end{multline} 
in which, on the right hand side, we have retained all the terms 
that are of the third degree in $\alpha_0$. Applying the binomial 
approximation for small $l^2$, and ignoring $\nu l^2$ and logarithmic 
variations because of their smallness, we get a solution, corrected to 
${\mathcal O} (l^2)$, as  
\begin{multline}
\label{kaynotcor} 
\beta \simeq \frac{1}{2} \ln \left(v_0\sqrt{gh_0}\right)  
\mp \int \frac{\nu}{2h_0^2\sqrt{gh_0}}\left(1 -  
\frac{3v_0}{v_0 \mp \sqrt{gh_0}}\right) \,{\mathrm d}r \\
\pm \int \frac{l^2 \omega^2 \Gamma_1 \Gamma_2 \sqrt{gh_0}}
{r(v_0 \mp \sqrt{gh_0})^3}\,{\mathrm d}r + c_2,
\end{multline} 
where $c_2$ is an integration constant. Applying the same
line of reasoning that followed Eq.~(\ref{kayminuscor}), we 
note that the third term on the right hand side of Eq.~(\ref{kaynotcor})
adds a surface tension-dependent correction, $\beta_l$, to $\beta_0$, 
in the full form, $\beta = \beta_0 + \beta_l$. And, as we have 
argued in the case of Eq.~(\ref{kayminuscor}), $l^2 \omega^2$ 
in Eq.~(\ref{kaynotcor}) renders 
the correction term subdominant to $\beta_0$, 
which is again consistent with our iterative treatment. 

From the wave solution in Eq.~(\ref{travsol}), which we now write  
as $f^\prime (r,t) = e^{-\beta}\exp (i\alpha -i\omega t)$, the 
amplitude part is extracted as 
$\vert f^\prime (r,t) \vert = e^{-\beta}$. 
Expressed in full, it is 
\begin{multline} 
\label{fampli} 
\big{\vert} f^\prime (r,t) \big{\vert} = \exp\left(-\beta \right) 
\sim \left(v_0 \sqrt{gh_0}\right)^{-1/2} \\
\times 
\exp\left[\pm \int \frac{\nu}{2h_0^2\sqrt{gh_0}}\left(1 -  
\frac{3v_0}{v_0 \mp \sqrt{gh_0}}\right) \,{\mathrm d}r\right] \\
\times
\exp\left[\mp \int \frac{l^2 \omega^2 \Gamma_1 \Gamma_2 \sqrt{gh_0}}
{r(v_0 \mp \sqrt{gh_0})^3}\,{\mathrm d}r \right].   
\end{multline} 
In the two exponential terms of Eq.~(\ref{fampli}), the upper signs 
correspond to a wave that propagates upstream against the radially 
outward flow of liquid. Such a wave undergoes a large blue-shift near 
the horizon as $v_0 \longrightarrow \sqrt{gh_0}$~\citep{su02}. 
The amplitude 
of the wave also suffers a large divergence here, because as the wave 
approaches the horizon through the subcritical region of the flow, 
where $v_0 < \sqrt{gh_0}$, both exponential terms in Eq.~(\ref{fampli}) 
diverge and ultimately result in 
$\vert f^\prime (r,t) \vert \longrightarrow \infty$. 
The exact opposite of this happens 
just inside the horizon, where, with $v_0 >\sqrt{gh_0}$,
the exponential terms in Eq.~(\ref{fampli}) vanish, resulting in 
$\vert f^\prime (r,t) \vert \longrightarrow 0$. 
Since these two completely contrasting features are segregated 
by the analogue 
event horizon, we regard the horizon as an impenetrable barrier 
--- a white hole ---
where the radial liquid outflow from the supercritical region blocks 
radially incoming waves from the subcritical region. 

Noting from Eq.~(\ref{radf}) that $f^\prime$ is a perturbation in
the volumetric flow rate, the divergence of $f^\prime$ implies an 
unstable pile-up of matter arbitrarily close to the horizon on the 
subcritical side, as theoretically expected about white holes in 
general~\citep{su02,dme74}, and as supported by laboratory 
experiments on circular hydraulic jumps~\citep{kdc07}.
Surface tension is more responsible than viscosity 
for the unstable pile-up 
because near the horizon the integral with surface tension in 
Eq.~(\ref{fampli}) has a singularity of the third order, whereas 
the integral with viscosity has a singularity of the first order. 
Hence, the divergence of $f^\prime$ is forced more by surface tension 
than by viscosity. While viscosity may cause a circular hydraulic 
jump to form at the horizon, surface tension is more effective 
than viscosity in 
blocking the passage of waves through the horizon, creating thus
a hydrodynamic white hole.  

\section{Penetrating the horizon} 
\label{sec6}
Waves propagating inwards against the steady radial outflow 
encounter a singularity at the horizon of the hydrodynamic 
white hole, where $v_0 = \sqrt{gh_0}$. This is obvious from the 
integrands in Eqs.~(\ref{kayminuscor}) and~(\ref{kaynotcor}).
In each case, circumventing the singularity requires
rendering it as a simple pole on the path of the integration,
and then applying Cauchy's residue theorem on the path~\citep{dk96}.
We first demonstrate this procedure for the simple
case of $\alpha = \alpha_0$ in Eq.~(\ref{kayminus}) 
by considering its upper sign, 
which stands for an inwardly travelling wave against the outflow. 
The main contribution to the integral comes from the immediate 
neighbourhood of $v_0 =\sqrt{gh_0}$, where $r=r_\star$, as 
follows from Eq.~(\ref{dhdr2}), leading up to Eq.~(\ref{rcrit}). 

A first-order Taylor expansion about the horizon gives
$v_0 -\sqrt{gh_0} \simeq (v_0 -\sqrt{gh_0})_{r_\star} 
+[{\mathrm d}(v_0 -\sqrt{gh_0})/{\mathrm d}r]_{r_\star}
(r-r_\star)$. The Taylor expansion in the neighbourhood of
the horizon transforms the singularity at $v_0 =\sqrt{gh_0}$
to a simple pole at $r=r_\star$. The zero-order term in the Taylor
expansion vanishes at the horizon, and with the first-order term 
we approximate Eq.~(\ref{kayminus}), for the upper sign, as
\begin{equation}
\label{alphahor} 
\alpha_0 \simeq 
\frac{\omega}{
\left[{\mathrm d}(v_0-\sqrt{gh_0})/{\mathrm d}r\right]_{r_\star}}
\int \frac{{\mathrm d}r}{r-r_\star}.
\end{equation}
At the white hole horizon, the analogue surface gravity~\citep{vis98} , 
\begin{equation} 
\label{surfgrav} 
G_\mathrm{s}=\sqrt{gh_0(r_\star)}
\left[\frac{\mathrm d}{{\mathrm d}r}
\left(\sqrt{gh_0} -v_0 \right)\right]_{r_\star}, 
\end{equation} 
and the analogue Hawking temperature~\citep{vis98},  
\begin{equation} 
\label{hawktemp} 
T_\mathrm{H}=\frac{\hbar G_\mathrm{s}}{2\pi k_\mathrm{B} 
\sqrt{gh_0(r_\star)}} = \frac{\hbar}{2\pi k_\mathrm{B}} 
\left[\frac{\mathrm d}{{\mathrm d}r}
\left(\sqrt{gh_0} -v_0 \right)\right]_{r_\star}.
\end{equation} 
In terms of $G_\mathrm{s}$ and
$T_\mathrm{H}$, the integral in Eq.~(\ref{alphahor}),
on extracting the residue at the pole, is reduced to  
\begin{equation} 
\label{alphres0} 
\alpha_0 \simeq 
-\frac{\hbar \omega \sqrt{gh_0(r_\star)}}{\hbar G_\mathrm{s}} 
\left(\pm i \pi \right)+{\mathcal P}\left[\alpha_0\right] =
-\frac{\hbar \omega}{2k_\mathrm{B}T_\mathrm{H}} 
\left(\pm i\right)+{\mathcal P}\left[\alpha_0\right], 
\end{equation}
where ${\mathcal P}[\alpha_0]$ is the
principal value of the integral. The negative sign in
$\pm i$ is due to a clockwise detour of
the pole, and the positive sign is due to an anti-clockwise
detour. Since both are mathematically valid, the choice
of the appropriate sign 
depends physically on the boundary condition at the pole~\citep{dk96}.

An additional contribution to $\alpha$ comes from 
surface tension, through the second term in Eq.~(\ref{kayminuscor}). 
For this term 
a first-order Taylor expansion gives
$(v_0 -\sqrt{gh_0})^4 \simeq (v_0 -\sqrt{gh_0})_{r_\star}^4 
+4(v_0 -\sqrt{gh_0})_{r_\star}^3
[{\mathrm d}(v_0 -\sqrt{gh_0})/{\mathrm d}r]_{r_\star}
(r-r_\star)$.
Now that we explicitly 
account for surface tension, we see from Eq.~(\ref{veesig})
that the zero-order term, $(v_0 -\sqrt{gh_0})_{r_\star}^4$, 
is ${\mathcal O} (l^2/r_\star^2)$ smaller that the first-order 
term in the Taylor expansion. Thus, we neglect the zero-order 
term and approximate
$(v_0 -\sqrt{gh_0})^4 \simeq  
4(v_0 -\sqrt{gh_0})_{r_\star}^3
[{\mathrm d}(v_0 -\sqrt{gh_0})/{\mathrm d}r]_{r_\star}
(r-r_\star)$.
This condition, imposed about the horizon, approximates the second 
term on the right hand side of Eq.~(\ref{kayminuscor}) (which we 
read as $\alpha_l$), with its upper sign, to
\begin{equation} 
\label{alphacorhor} 
\alpha_l \simeq  
\left[\frac{l^2 \omega^3 \Gamma_1 \sqrt{gh_0}}
{8(v_0 - \sqrt{gh_0})^3 \,
{\mathrm d}(v_0-\sqrt{gh_0})/{\mathrm d}r}\right]_{r_\star}
\int \frac{{\mathrm d}r}{r-r_\star}.
\end{equation}
The wave number, $\kappa (r)= 2\pi/\lambda(r) =\omega/(v_0-\sqrt{gh_0})$, 
using which we define a relevant frequency in the system, 
\begin{equation} 
\label{freqomeg} 
\Omega = \frac{l^2 \omega^3 \Gamma_1 \sqrt{gh_0}}
{8(v_0 -\sqrt{gh_0})^3}
= \frac{l^2\kappa^3 \Gamma_1 \sqrt{gh_0}}{8}
= \frac{\pi^3 \Gamma_1 \sqrt{gh_0}}{\lambda}
\left(\frac{l}{\lambda}\right)^2.
\end{equation} 
With the definitions in Eq.~(\ref{freqomeg}), 
along with Eqs.~(\ref{surfgrav}) and~(\ref{hawktemp}), 
we evaluate the integral in Eq.~(\ref{alphacorhor}) to be 
\begin{equation} 
\label{alphresl} 
\alpha_l \simeq -\left[
\frac{l^2 \kappa^3 \Gamma_1 gh_0}{8G_\mathrm{s}} \right]_{r_\star}
\left(\pm i \pi \right)+{\mathcal P}\left[\alpha_l\right] 
= -\frac{\hbar \Omega}{2k_\mathrm{B}T_\mathrm{H}} 
\left(\pm i\right)+{\mathcal P}\left[\alpha_l\right], 
\end{equation}
with ${\mathcal P}\left[\alpha_l\right]$ being the principal value 
of the integral. The implication of either 
sign in $\pm i$ adheres to the same 
principle as has been discussed following Eq.~(\ref{alphres0}). 

The result in Eq.~(\ref{alphres0}) relates to the flow condition
in which surface tension has not been considered, as discussed
in Sec.~\ref{sec3sub2}. In this case, the steady liquid outflow
has an inner radial solution and an outer radial solution, which 
are joined discontinuously at the circular hydraulic jump. Now, 
the jump visibly coincides with the horizon of the hydrodynamic 
white hole, which stands as an unyielding barrier to waves that 
approach it from the subcritical region of the flow. Therefore, for
waves that travel upstream in the subcritical region, the horizon 
is a rigid inner boundary, where the waves are blocked, 
piled up and compressed. This is generally expected for both 
general relativistic white
holes~\citep{dme74} and their fluid analogues~\citep{su02}.
However, notwithstanding the rigidity of this boundary,
Eq.~(\ref{alphres0}) suggests that a wave may yet penetrate the 
horizon. This effect is enhanced by surface 
tension, which, as shown in Eqs.~(\ref{veesig}) and~(\ref{rsig}),   
softens the horizon, thus making it more penetrable. 
This is what Eq.~(\ref{alphresl}) shows. 

The wave solution in Eq.~(\ref{travsol}) has $s(r)=\alpha (r)+i\beta (r)$
and, further, $\alpha (r) = \alpha_0 (r)+ \alpha_l (r)$, 
as in Eq.~(\ref{kayminuscor}).
Hence, the amplitude of the wave that penetrates the horizon,
$\vert f_\mathrm{P}^\prime \vert$, is determined by 
$\Im(\alpha_0)$ in Eq.~(\ref{alphres0}) and
$\Im(\alpha_l)$ in Eq.~(\ref{alphresl}). Together they give  
\begin{equation} 
\label{amplipen} 
\big{\vert} f_\mathrm{P}^\prime \big{\vert}
\sim \exp \left[ \pm
\frac{\hbar (\omega + \Omega)}{2k_\mathrm{B}T_\mathrm{H}}\right]. 
\end{equation} 
The horizon stands as a strong barrier against an incoming wave 
that travels upstream, counter to the fluid outflow,  
and tries to enter the supercritical region 
from the subcritical region. This 
being the physical boundary
condition at the horizon, a wave can only tunnel through       
it with a decaying amplitude. The tunnelling 
amplitude thus corresponds to the negative 
sign in Eq.~(\ref{amplipen}), with the tunnelling probability 
given by $\vert f_\mathrm{P}^\prime \vert^2$. 
Surface tension plays a crucial part in the tunnelling 
because the frequency, $\Omega$, is 
set in terms of surface tension, as Eq.~(\ref{freqomeg}) shows. 
What is more, in the tunnelling amplitude, $\hbar \Omega$
is scaled by the fluid analogue of the Hawking temperature, 
$T_\mathrm{H}$, which makes the tunnelling phenomenon a case of 
Hawking radiation in fluid 
analogues~\citep{wgu81,tj91,wgu95,vis98,su02,blv11}.
The combined outcome of these two facts
is that surface tension becomes the most obvious physical means 
by which Hawking quanta 
penetrate the horizon of the hydrodynamic white hole.

Surface tension also prevents the incoming waves from undergoing 
an arbitrarily high blue-shift near the horizon. This is clear 
from Eq.~(\ref{phasegrav}), which shows that if $l=0$, then 
$v_{\mathrm g} \longrightarrow 0$ at the horizon for a radially 
convergent wave packet travelling against the outflowing fluid. 
The wavenumber, $\kappa$, will consequently be blue-shifted without 
limit near the horizon, and the corresponding wavelenghth, $\lambda$, 
will be shortened arbitrarily. These difficulties have been known 
for long with regard to the fluid analogues of Hawking radiation, 
and they have been addressed
variously~\citep{tj91,wgu95,bmps95,cj96}.
However, when surface tension is accounted for, i.e. 
when $l \neq 0$, the physical 
conditions become qualitatively different.
From Eqs.~(\ref{veesig}),~(\ref{rsig}) and~(\ref{phasegrav}), 
we realize that surface tension creates a thin layer 
of uncertainty about the exact horizon conditions that result from 
Eq.~(\ref{dhdr2}), namely, $r = r_\star$ and $v =\sqrt{gh_\star}$. 
The relative thickness of this layer is 
${\mathcal O} (l^2/r_\star^2)$, which, though a small fraction, 
is still enough to restrict  
$v_{\mathrm g}$ to a small non-zero value
(instead of just vanishing) near the horizon. Thus, close to 
the horizon the blue-shifting is limited by the capillary length, 
a condition that can only be attributed physically to surface tension.
By this then an incoming wave packet can avoid an infinite 
blue-shift near the horizon and can tunnel through the thin 
zone of uncertainty about the white-hole barrier.\footnote{Viscous 
dissipation in 
spherically symmetric transonic astrophysical accretion 
similarly enables
Hawking phonons to tunnel through the horizon of an acoustic black 
hole at the sonic radius of the inflowing gas~\citep{akr20}.}

While tunnelling is associated with the negative sign on the 
right hand side of Eq.~(\ref{amplipen}), we also 
consider the consequence of a wave that penetrates the horizon 
with a positive sign in the amplitude. Such a wave will cause an 
instability 
about the horizon of the hydrodynamic white hole. Since surface 
tension is the essential physical factor in the amplitude of the 
penetrating wave, we note that large surface tension can destabilize
the steady hydraulic jump and even make it disappear~\citep{kas08}. 
We compare this behaviour of a hydrodynamic white hole with 
what happens in a general relativistic white hole. In the latter 
context white holes force a pile-up at the 
horizon, resulting in a strong blue-shift~\citep{dme74}. 
Instabilities arise in consequence and cause white holes to
disappear~\cite{dme74,ws80}. 

\section{Concluding remarks}
\label{sec7}
Since detecting Hawking radiation through direct observations 
of astrophysical systems is not very likely to succeed, 
it becomes necessary instead to use fluid analogues of gravity.
The fluid in question can be a gas or a liquid. Either
form of matter will bring forth its own set of physical properties, 
with qualitatively different outcomes. For instance, in a radially
converging gas flow, as in spherically symmetric astrophysical
accretion, surface tension is not relevant in the Hawking process 
but viscosity (even weak molecular viscosity) facilitates 
it~\cite{akr20}. In contrast, our present study, involving 
a free-surface shallow liquid (water) flow, shows that surface 
tension is the chief player in the Hawking process that happens 
at the position of the hydraulic jump. Hence, experiments that 
use fluid analogues of gravity  
can exploit the various physical attributes of fluids 
to test gravitational theories.

The base state in our study is dominated by gravity (the 
long-wavelength regime) in which viscosity brings about the 
discontinuity of the hydraulic jump at the location of the 
hydrodynamic horizon. Surface tension is introduced as a 
small effect in this base state, as guided by the criterion 
that the capillary length is far less than the jump radius.     
As a result, surface tension has a significant influence only
around the hydraulic jump, where the free-surface
height of the flow has a large gradient. The jump is formed
because of viscosity and its location is scaled by gravity.  
The main effect of surface tension here is in the pile-up
and in the horizon penetration. However, qualitatively 
different conditions will obtain when the jump radius is 
comparable to the capillary length, as happens in 
superfluids~\citep{rgp07} and metal femtocups~\citep{rama07}.
Capillary effects are dominant in these cases. This point 
may be relevant to fluid analogues of gravity as well.  
We have shown how capillary length creates a thin
layer of uncertainty about the circular radius of the jump. 
This fluid analogue can be compared with the Planck length 
about the Schwarzschild radius of a general relativistic 
black hole~\citep{swh75}. Now, for a black 
hole of Planck mass both length scales will be the same. 
A possible fluid analogue of this could then be a hydraulic 
jump whose radius is of the order of the capillary length.

\begin{acknowledgments}
The authors thank J. K. Bhattacharjee for useful comments.  
\end{acknowledgments}

\bibliography{jr2025} 
\end{document}